\documentclass[prl,superscriptaddress,twocolumn,nofootinbib,showpacs,preprintnumbers,floatfix]{revtex4-2}

\usepackage{mathrsfs}
\usepackage{amsfonts,amssymb,amsmath}
\usepackage{graphicx,epsfig}
\usepackage{multirow}
\usepackage{verbatim}
\usepackage{color}

\usepackage{slashed}

\declareslashed{}{/}{0}{-0.05}{E}
\declareslashed{}{/}{0}{-0.05}{P}
\declareslashed{}{/}{0}{-0.10}{\Widehat{P}}

\def\fsu5{$\cal{F}$-$SU(5)$}
\def\bfsu5{$\boldsymbol{\mathcal{F}}$-$\boldsymbol{SU(5)}$}

\def\mv{$M_{\rm V}$}

\def\lsp{$\widetilde{\chi}_1^0$}
\def\gl{$\widetilde{g}$}
\def\ls{$\widetilde{t}_1$}
\def\char{$\widetilde{\chi}_1^{\pm}$}
\def\sel{$\widetilde{e}_R$}
\def\neu2{$\widetilde{\chi}_2^0$}
\def\stau{$\widetilde{\tau}^{\pm}_1$}
\def\m12{$M_{1/2}$}
\def\m3half{$M_{3/2}$}
\def\m32{$M_{32}$}

\def\fb{${\rm fb}^{-1}$~}

\def\bs0{$B_S^0 \rightarrow \mu^+ \mu^-$}

\def\bea{\begin{eqnarray}}
\def\eea{\end{eqnarray}}

\begin{document}

\title{Spinning No-Scale $\cal{F}$-$SU$(5) in the Right Direction}

\author{Tianjun Li}

\affiliation{CAS Key Laboratory of Theoretical Physics, Institute of Theoretical Physics, 
Chinese Academy of Sciences, Beijing 100190, P. R. China}

\affiliation{ School of Physical Sciences, University of Chinese Academy of Sciences, 
No.19A Yuquan Road, Beijing 100049, P. R. China}

\author{James A. Maxin}

\affiliation{Department of Chemistry and Physics, Louisiana State University, Shreveport, Louisiana 71115 USA}

\author{Dimitri V. Nanopoulos}

\affiliation{George P. and Cynthia W. Mitchell Institute for Fundamental Physics and Astronomy, Texas A$\&$M University, College Station, TX 77843, USA}

\affiliation{Astroparticle Physics Group, Houston Advanced Research Center (HARC), Mitchell Campus, Woodlands, TX 77381, USA}

\affiliation{Academy of Athens, Division of Natural Sciences, 28 Panepistimiou Avenue, Athens 10679, Greece}


\begin{abstract}

The Fermi National Accelerator Laboratory (FNAL) recently announced confirmation of the Brookhaven National Lab (BNL) measurements of the $g-2$ of the muon that uncovered a discrepancy with the theoretically calculated Standard Model value. We suggest an explanation for the combined BNL+FNAL 4.2$\sigma$ deviation within the supersymmetric grand unification theory (GUT) model No-Scale \fsu5 supplemented with a string derived TeV-scale extra $10+\overline{10}$ vector-like multiplet and charged vector-like singlet $(XE,XE^c)$, dubbed flippons. We introduced these vector-like particles into No-Scale Flipped $SU(5)$ many years ago, and as a result, the renormalization group equation (RGE) running was immediately shaped to produce a distinctive and rather beneficial two-stage gauge coupling unification process to avoid the Landau pole and lift unification to the string scale, in addition to contributing through 1-loop to the light Higgs boson mass. The flippons have long stood ready to tackle another challenge, and now do so yet again, where the charged vector-like ``lepton''/singlet couples with the muon, the supersymmetric down-type Higgs $H_d$, and a singlet $S$, using a chirality flip to easily accommodate the muonic $g-2$ discrepancy in No-Scale \fsu5. Considering the phenomenological success of this string derived model over the prior 11 years that remains accommodative of all presently available LHC limits plus all other experimental constraints, including no fine-tuning, and the fact that for the first time a Starobinsky-like inflationary model consistent with all cosmological data was derived from superstring theory in No-Scale Flipped $SU(5)$, we believe it is imperative to reconcile the BNL+FNAL developments within the model space.

\end{abstract}



\preprint{ACT-02-21, MI-HET-753}

\maketitle


\section{Introduction}

The anomalous magnetic moment of the muon $g-2$ has long been envisioned as a doorway to infiltrate beyond the Standard Model (BSM) physics. In the Standard Model, loop corrections deviate from the tree-level $g = 2$ gyromagnetic ratio, leading to an anomalous magnetic moment expressed in terms of $a_{\mu} \equiv (g_{\mu} - 2)/2$. Given the precision with which the magnetic moment of the muon can be measured and the sensitivity of the muon's magnetic moment to new physics, tremendous energy has been expended to calculate the Standard Model's contribution $a_{\mu}(\rm SM)$~\cite{Davier:2010nc,davier:2017zfy,keshavarzi:2018mgv,colangelo:2018mtw,hoferichter:2019gzf,davier:2019can,keshavarzi:2019abf,kurz:2014wya,gerardin:2019rua,Aubin:2019usy,giusti:2019hkz,melnikov:2003xd,masjuan:2017tvw,Colangelo:2017fiz,hoferichter:2018kwz,gerardin:2019vio,bijnens:2019ghy,colangelo:2019uex,pauk:2014rta,danilkin:2016hnh,jegerlehner:2017gek,knecht:2018sci,eichmann:2019bqf,roig:2019reh,Blum:2019ugy,colangelo:2014qya,Aoyama:2012wk,Aoyama:2019ryr,czarnecki:2002nt,gnendiger:2013pva}, up to 10th order in QED (For example, see Ref.~\cite{Aoyama:2012wk}). Experiments can then assess the difference $\Delta a_{\mu} = a_{\mu}(\rm Exp) - a_{\mu}(\rm SM)$ between the theoretical and observed values as a probe into unknown contributions to the anomalous magnetic moment.

The first precision measurements of $a_{\mu}$ were undertaken by the Brookhaven National Lab (BNL)~\cite{bennett:2006fi}, uncovering a 3.7$\sigma$ discrepancy with the Standard Model prediction. The BNL variance certainly induced intrigue, but the large error bars left room for skepticism of its longevity and conceded the plausible conclusion that only a statistical fluctuation had surfaced. The Fermi National Accelerator Laboratory (FNAL) launched an independent venture to measure $a_{\mu}$ in an effort to confirm the original BNL findings and phase out all ambiguity. The FNAL version of the $g-2$ experiment steadily accumulated data over several years, ultimately announcing recently an updated figure of~\cite{Abi:2021gix}
\begin{equation}
\label{muon} \Delta a_\mu = a_\mu(\rm Exp) - a_\mu(SM) = (25.1\pm5.9)\times 10^{-10}~,~\,
\end{equation}
and just as material, a substantially reduced factor of uncertainty. In toto, the combined BNL+FNAL result unmasks a statistically significant 4.2$\sigma$ deviation from the Standard Model calculation. What cannot be dismissed is that the magnitude of both the BNL and FNAL $\Delta a_{\mu}$ are of comparable scale to the electroweak contribution to $a_{\mu}(\rm SM)$~\cite{czarnecki:2002nt,gnendiger:2013pva}, and this strongly suggests pervasive new TeV-scale physics.

The dearth of any historic BSM discoveries thus far notwithstanding, supersymmetry (SUSY) perseveres as the most promising extension to the Standard Model. Besides offering a natural method to ensure quantum stability of the scalar Higgs field, SUSY further guarantees gauge coupling unification, provides a dark matter candidate in the form of the lightest supersymmetric particle (LSP) under R-parity, and imposes a much needed radiative electroweak symmetry breaking mechanism. Perhaps first and foremost though, SUSY constructions are organic within superstring theory. While recorded data of proton-proton collisions at the Large Hadron Collider (LHC) Run 1 and Run 2 have been barren of SUSY candidate events, experiments such as $g-2$ of the muon engineer a parallel route to probing the elusive SUSY.

The phenomenology of the supersymmetric grand unified theory (GUT) model No-Scale \fsu5 has been well studied and documented (for example, see Refs.~\cite{Li:2010rz,Li:2010ws,Li:2010mi,Maxin:2011hy,Li:2011xua,Li:2011in,Li:2011gh,Li:2011ab,Li:2013naa,Li:2013mwa,Li:2016bww,Ford:2019kzv}). Tantamount to these compelling phenomenological developments, recent cosmological work demonstrated the derivation of a complete Starobinsky-like inflationary model from superstring theory within No-Scale Flipped $SU(5)$, all while also generating right-handed neutrino masses that can alleviate the baryon asymmetry of our universe~\cite{Antoniadis:2020txn}, described as a $``\lambda_6~{\rm Universe}"$ (for instance, see Refs.~\cite{Ellis:2019jha,Ellis:2019bmm,Ellis:2019hps,Ellis:2019opr,Ellis:2020qad,Ellis:2020xmk,Ellis:2020nnp,Nanopoulos:2020nnh,Ellis:2020krl}~\cite{Antoniadis:2020txn}). Indeed, the phenomenological strength of No-Scale \fsu5 promptly asserted in 2010 the model would be a prominent candidate for $the$ natural GUT of the universe we occupy. In these prior phenomenological works, the model complied with all BSM experiments, the one notable exception being the SUSY contribution to the anomalous magnetic moment of the muon. Neglecting the contribution of additional vector-like multiplets introduced into No-Scale \fsu5 from its outset, the computed SUSY contribution persisted near $\Delta a_{\mu}(\rm SUSY) \sim 1 \times 10^{-10}$~\cite{Li:2011xua,Li:2013naa,Li:2016bww,Ford:2019kzv}. Whereas this was consistent with the 3$\sigma$ intervals $-17.1 \times 10^{-10} \le \Delta a_{\mu} \le 43.8 \times 10^{-10}$ obtained from the original BNL experiment~\cite{bennett:2006fi}, the tension with the BNL central value was nonetheless worrisome. Despite the model strain from inconsistency with the measured $g-2$ of the muon, the schism was always burdened with the crucial qualification that no contributions from the vector-like multiplets had ever been integrated with the SUSY computation. Hence, the SUSY only allotment had always been believed to be shy of what \fsu5 could deliver in total. Fast forwarding to present day, the announcement that FNAL has confirmed the original BNL findings while mutually reducing the uncertainty by a factor of four, the need to reconcile this tension is now undeniable. 

Our solution to the diverging muon $g-2$ value in No-Scale \fsu5 by embracing the vector-like enhancement will be presented later in this paper. First, we shall endeavor to summarize the primary elements and salient achievements of the model, then close with a survey of the model's prevailing standing with regards to the LHC Run 2  SUSY constraints.

\section{No-Scale \bfsu5}

Supersymmetric models motivated by string model building tactics that are testable at the LHC Run 2 linger as an investigation of fundamental importance. One well-developed model fulfilling this stipulation is No-Scale \fsu5, owing its inception to the goal of building and duly substantiating a realized vacuum of F-Theory. No-Scale \fsu5 adheres to rigid top-down theoretical dynamics concurrent with rigorous phenomenological bottom-up constraints, where continuity amidst these two scales is non-trivial. Fittingly, the observation of explicit correlations between top-down and bottom-up precepts in \fsu5 has proven to be quite meaningful.

\begin{figure}[t]
   \centering
        \includegraphics[width=0.45\textwidth]{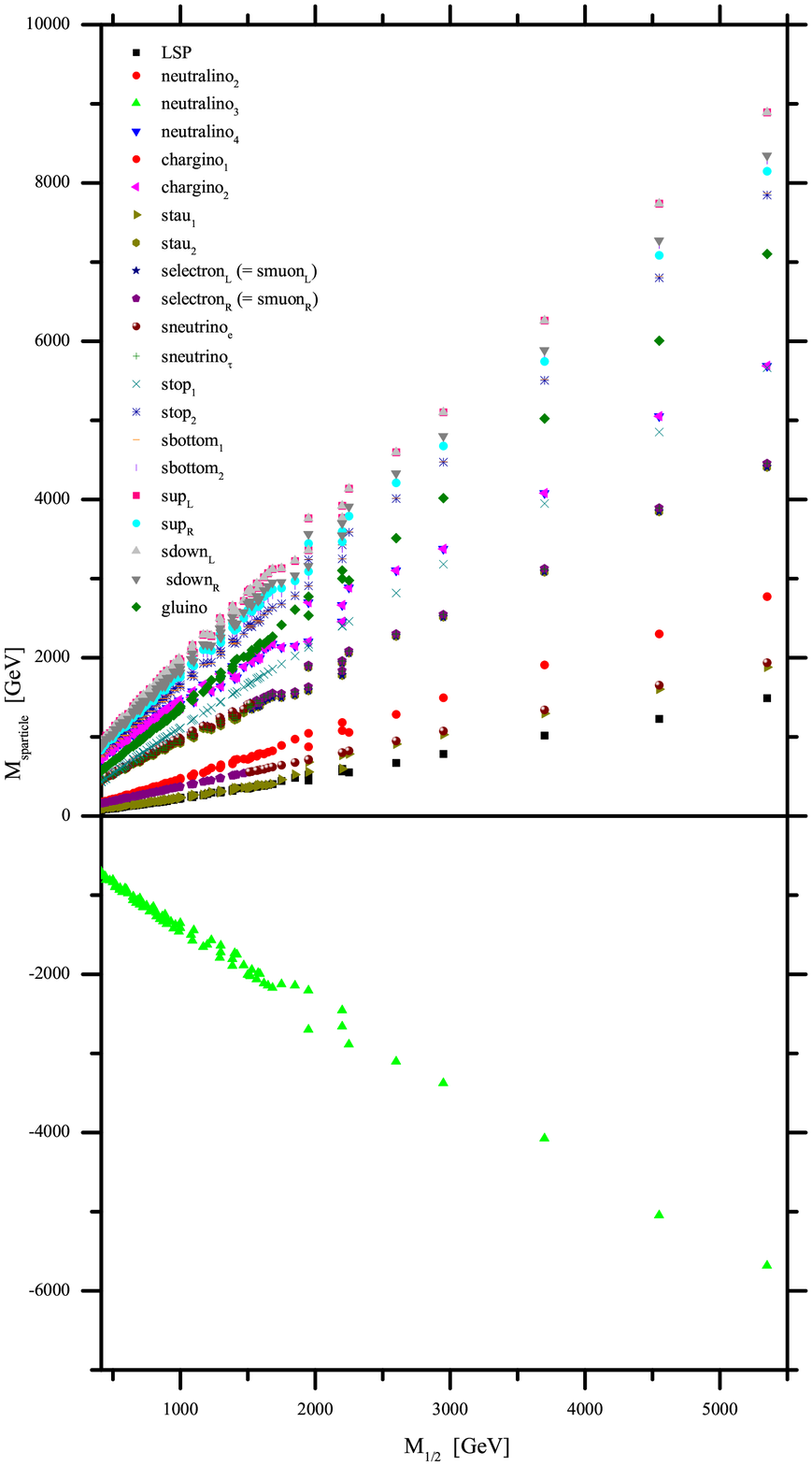} \\
        \caption{Defining relationship amongst the sole model parameter $M_{1/2}$ and the SUSY particles. The linear dependence of all SUSY masses on $M_{1/2}$ emerging from No-Scale SUGRA is clear. Only actual computed points are plot, borrowed from all the benchmark sets published in Refs.~\cite{Li:2011xua,Li:2011in,Li:2011gh,Li:2013naa,Li:2013mwa,Li:2016bww,Ford:2019kzv}.}
        \label{fig:masses}
\end{figure}

Upon the requisite localization demanded within string theory, SUSY transforms into supergravity (SUGRA). Introduction of the SUGRA framework is critical, but further elements are demanded, for instance, the longevity and cosmological flatness of our universe, which are very desirable theoretically though rather non-trivial in practice. No-Scale SUGRA~\cite{Cremmer:1983bf,Ellis:1983sf, Ellis:1983ei, Ellis:1984bm, Lahanas:1986uc} answers the call here, facilitating a SUSY breaking mechanism that guarantees the cosmological constant vanishes at tree-level, spurring the aforementioned longevity and flatness~\cite{Cremmer:1983bf}. Additionally, No-Scale SUGRA empowers a dynamic determination mechanism of the gravitino mass $M_{3/2}$, which in parallel dynamically determines the unified gaugino mass $M_{1/2}$ through their proportional equivalence. In the most elementary No-Scale SUGRA, the SUSY breaking scale is $M_{1/2}$, and consequently all SUSY masses are effectively determined, as well as the $Z$-boson and top quark masses. The linear relationship between all SUSY masses and $M_{1/2}$ is vividly displayed in FIG.~\ref{fig:masses}, where only actual computed points are plot, borrowed from all the benchmark sets published in Refs.~\cite{Li:2011xua,Li:2011in,Li:2011gh,Li:2013naa,Li:2013mwa,Li:2016bww,Ford:2019kzv}. This FIG.~\ref{fig:masses} will be significant in the fine-tuning discussion. The dependence of all model scales upon $M_{1/2}$ affirms that No-Scale \fsu5 is a bone fide natural one-parameter model.

The prospect of gauge unification would firmly imply the influence of a GUT. The minimal supersymmetric $SU(5)$ GUTs are afflicted with anomalous doublet-triplet splitting and dimension-5 proton decay emerging from colored higgsino exchange~\cite{Antoniadis:1987dx}. The Flipped $SU(5)$ GUT model~\cite{Barr:1981qv,Derendinger:1983aj,Antoniadis:1987dx,Nanopoulos:2002qk} conquers these aberrations through the missing partner mechanism~\cite{Antoniadis:1987dx}, and as such, promotes natural doublet-triplet splitting and suppression of dimension-5 proton decay. Flipped $SU(5)$ also furnishes fundamental GUT-scale Higgs representations (not adjoints), a natural nullification of CP violation and flavour changing neutral currents, and a two-step see-saw mechanism for neutrino masses~\cite{Ellis:1992nq,Ellis:1993ks}.

The defining Kahler potential~\cite{Ellis:1984bm} invoked in No-Scale SUGRA compels that $M_0$, $A_0$, and $B_{\mu}$ are uniformly zero at the $M_{\cal F}$ boundary, representing the ultimate $SU(5) \times U(1)_X$ unification scale. By virtue of the renormalization group equations (RGEs), all soft SUSY breaking terms evolve down from the exclusive high-energy mass parameter $M_{1/2}$. Dual sets of TeV-scale vector-like multiplets derived from local F-Theory model building ~\cite{Jiang:2006hf} enter the RGE evolution, which we tagged with the vivid name of $flippons$~\cite{Li:2010mi}. The merging of Flipped $SU(5)$ with the vector-like multiplets extracted from F-Theory is designated \fsu5~\cite{Jiang:2008yf, Jiang:2009za}. Evasion of a landau pole for the strong coupling constant restrains the vector-like multiplets to only two particular sets~\cite{Jiang:2006hf}. The vector-like multiplets shift the RGE $\beta$-coefficients, successively boosting the $SU(5) \times U(1)_X$ unification to the $M_{\cal F}$ scale~\cite{Jiang:2006hf, Jiang:2008yf, Jiang:2009za}~\cite{Li:2010ws}, within close proximity of the string and Planck scales. Jointly, a second stage unification $SU(3)_C \times SU(2)_L$ ensues at the scale $M_{32}$, near the traditional MSSM GUT scale. Emanating out of this split unification operation is a flat $SU(3)$ RGE running (due to $b_3 = 0$ enforced by the vector-like multiplets) from $M_{32}$ down to the TeV scale at which all the vector-like multiplets decouple, referred to as the scale $M_V$. The mass scale $M_V$ is reserved as a free model parameter, though phenomenologically viable results reveal $M_V$ is restricted to only a rudimentary function of $M_{1/2}$, increased evidence of the deep correspondence intrinsically cemented within all mass scales of No-Scale \fsu5.

\section{String Derived Inflationary Model}

Recently, the No-Scale Flipped $SU(5)$ framework has been directly derived from superstring theory. Indeed, using the free-fermionic formulation (fff) of superstring theory provided a Starobinsky-like inflationary model based on No-Scale Flipped $SU(5)$~\cite{Antoniadis:2020txn}. To our knowledge, this is the first time that a full-fledged successful inflationary model has emerged from superstring theory. The derivation involves computation of higher orders in $\alpha'$, until 8th order, and the relevant superpotential coefficients were not only calculable but right! For example, the magnitude of the inflationary scale, $\frac{M_I}{M_{St}} \sim 10^{-5}$, as well as the ratio of the trilinear term $\frac{\lambda}{M_I} \sim {\cal O}(1)$, are consequences of the larger order correction in $\alpha'$. Amazingly though, there is another deep connection between inflation and neutrino masses. The inflaton field, $\phi$, is identified as the $SU(5) \times U(1)$ singlet that couples to $F_n$, which is the $\underline{10}$ coupling the top quark and $\overline{F}$, where $\overline{F}$ is the $\overline{10}$ Higgs field that is tied together with the $F_1$, given that $F_1$ is the 10 Higgs field that breaks down the $SU(5) \times U(1)$ initially to $SU(3) \times SU(2) \times U(1)$, $i.e.$, the $\lambda_6 F_n \overline{F} \phi$ coupling plays an ubiquitous role in our theory. It provides the decay modes of the inflaton, thus reheating, and when the $\overline{F}$ gets a vacuum expectation value (VEV) it provides right-handed neutrino masses, through which we can get the baryon asymmetry of the universe, throughout the recycling of lepton asymmetry, via non-perturbative effects of the EW scale. Notice that the inflaton is the bosonic part in the Q-superfield, where the fermionic part is the right-handed neutrino. Therefore, $M_{inflaton} \approx M_{Rh-\nu}$, assuming $M_{SUSY} \approx {\cal O}$(TeV) scale, hence we understand for the first time why the mass of the inflaton is the same as the right-handed neutrino mass, since it is part of the two component of the same chiral superfield $\phi$! For a review of this $``\lambda_6~{\rm Universe}"$ framework see Refs.~\cite{Ellis:2019jha,Ellis:2019bmm,Ellis:2019hps,Ellis:2019opr,Ellis:2020qad,Ellis:2020xmk,Ellis:2020nnp,Nanopoulos:2020nnh,Ellis:2020krl}~\cite{Antoniadis:2020txn}, thus we feel justified to try to use such a framework to see if we can accommodate the 4.2$\sigma$ discrepancy between experiment and theory in the recent combined BNL+FNAL results.

\section{No Fine-tuning}

Electroweak fine-tuning persists as the obstacle that plagues all supersymmetric GUT models. Natural SUSY, also known as naturalness, demands that a solution to the Higgs potential tree-level minimization proceed with no disproportionate cancellations. The minimization with respect to the $H_u$ and $H_d$ directions is

\begin{eqnarray}
\frac{M_Z^2}{2} = \frac{M_{H_d}^2  - \tan^2\beta ~M_{H_u}^2}{\tan^2\beta -1} -\mu^2~,~
\label{eq:ewmin}
\end{eqnarray}

\noindent though the effective scalar potential $V_{\rm eff} \to V_{\rm tree} + V_{\rm loop}$ must be injected with the radiative corrections $M^2_{H_u} \to M^2_{H_u} + \Sigma_u^u$ and $M^2_{H_d} \to M^2_{H_d} + \Sigma_d^d$, the majority of which arise from the top squarks $\widetilde{t}_1$ and $\widetilde{t}_2$. The dilemma here is clear. The freedom on $M_{H_u}$, $M_{H_d}$, $M_{\widetilde{t}_1}$, $M_{\widetilde{t}_2}$, $\mu$, and tan$\beta$ is rather advantageous for uncovering a solution to Eq.~(\ref{eq:ewmin}), given $M_Z$ is a known empirical quantity. Nonetheless, this freedom disguises the unsavory fact that these terms are being unnaturally tuned to deliver a viable solution. The remedy for eliminating all fine-tuning regrettably bears the scars of an all too often empty cupboard of solution sets, thereby rendering a model impotent.

Here is where the overall leading order rescaling in terms of the mass scale $M_{1/2}$ and No-Scale SUGRA is critical, as illustrated in FIG.~\ref{fig:masses} using all the benchmark sets published in Refs.~\cite{Li:2011xua,Li:2011in,Li:2011gh,Li:2013naa,Li:2013mwa,Li:2016bww,Ford:2019kzv}. The proportional dependence of all SUSY mass parameters upon this single unified mass term $M_{1/2}$ prevents any explicit dialing up of desirable solutions right from the start. Moreover, the strict No-Scale condition $B_{\mu}$($M_{\cal F}$) = 0 in effect determines the value of tan$\beta$ at low-energy. Let's face it: It is impossible to willfully tune $M_{H_u}$, $M_{H_d}$, $M_{\widetilde{t}_1}$, $M_{\widetilde{t}_2}$, $\mu$, and tan$\beta$ in No-Scale \fsu5 when there is only one degree of freedom in the entire model, and thus all mass parameters are functions of the sole model variable $M_{1/2}$. In essence, $M_{H_u}$, $M_{H_d}$, $M_{\widetilde{t}_1}$, $M_{\widetilde{t}_2}$, $\mu$, and tan$\beta$ are already effectively determined once a numerical value for $M_{1/2}$ is chosen and the No-Scale boundary conditions are set. This is built into the No-Scale \fsu5 model due to the merging of No-Scale SUGRA with Flipped $SU(5)$. No-Scale SUGRA essentially acts as a natural safeguard against fine-tuning and indeed guarantees no fine-tuning can ever materialize.

The absence of fine-tuning can be analytically shown as well~\cite{Leggett:2014mza,Leggett:2014hha}. A methodology to numerically measure fine-tuning was first introduced 35 years ago by Ellis, Enquist, Nanopoulos, and Zwirner (EENZ)~\cite{Ellis:1986yg, Barbieri:1987fn}:

\begin{equation}
\Delta_{\rm EENZ} \equiv \left| \frac{\partial \ln (M_Z)}{\partial \ln (\widetilde{m}_i)}\right| \;,
\label{eq:delta}
\end{equation}

\noindent where $M_Z$ is the $Z$-boson pole mass and $\widetilde{m}_i$ represent the mass terms in Eq.~(\ref{eq:ewmin}). Following upon the discussion of the prior paragraph, $M_Z$ is trivially a function of $M_{1/2}$. We fix the $Z$-boson mass at its experimental value in our numerical computations, though the variation of $M_Z$ with respect to $M_{1/2}$ is effectively witnessed though variations in the ratio between the $\mu$ parameter and $M_{1/2}$~\cite{Leggett:2014hha}. This important relationship between $M_Z$ and $M_{1/2}$ can be described by

\begin{eqnarray}
M_Z^n ~\simeq~ f_n \left( c_i \right) ~M_{1/2}^n~,~\,
\label{eq:fn}
\end{eqnarray}

\noindent for all orders of $n$ (regarding our purposes here, most importantly for $n$ = 1,2). Note that given the linear alliance springing from the rescaling property of $M_{1/2}$, $f_n$ is a constant dimensionless parameter that is itself a function of the $c_i$. The $c_i$ designate additional dimensionless parameters, for instance, the gauge and Yukawa couplings, together with the ratios between $\mu$ and $M_{1/2}$ and between $M_V$ and $M_{1/2}$, etc. Fine-tuning can be measured by exposing the maximum $\Delta_{EENZ}$ for all $\widetilde{m}_i$, however, as noted, there is only one degree of freedom, $M_{1/2}$, to investigate. From Eq.~(\ref{eq:fn}) we have

\begin{eqnarray}
\frac{\partial M_Z^n}{\partial M_{1/2}^n} ~\simeq~ f_n~,~\, 
\label{eq:fnd}
\end{eqnarray}

\noindent and inserting Eq.~(\ref{eq:fn}) and Eq.~(\ref{eq:fnd}) into Eq.~(\ref{eq:delta}) we then arrive at the significant result of

\begin{eqnarray}
\Delta_{EENZ} = \left| \frac{\partial{\rm ln}(M_Z^n)}{\partial {\rm ln}(M_{1/2}^n)} \right| =  \frac{M_{1/2}^n}{M_Z^n} \frac{\partial M_Z^n}{\partial M_{1/2}^n} \simeq \frac{1}{f_n} f_n = 1
\label{eq:fnpar}
\end{eqnarray}

\noindent for all $n$, and we therefore find that $\Delta_{EENZ} \simeq 1$ and we are thus assured of no fine-tuning.

\section{Phenomenology}

Beyond the discrepant $\Delta a_{\mu}$ already reported, \fsu5 remains consistent with all other experimental constraints. The observation of a light Higgs boson mass of $M_h = 125.1 \pm 0.14$~GeV~\cite{Aad:2012tfa,Chatrchyan:2012xdj} wrecked havoc upon the SUSY GUT model space in 2012, though \fsu5 could easily elevate the Higgs boson mass to 125~GeV though the vector-like multiplets coupling to the Higgs fields via the vector-like particle Yukawa coupling~\cite{Li:2011ab}. This was essential in the low mass region of M(\gl) $\lesssim$ 1.5~TeV, where we implemented the maximum Yukawa coupling, though for M(\gl) $\gtrsim$ 1.5~TeV, the 2-loop coupling to the heavier light stops via the top Yukawa coupling was sufficient to raise the light Higgs mass to its measured value and therefore the minimum vector-like particle Yukawa coupling was applied~\cite{Li:2016bww,Ford:2019kzv}. On the contrary, for the \fsu5 D-brane model~\cite{DeBenedetti:2018fxa,DeBenedetti:2019hrk}, which does not require the vanishing No-Scale boundary conditions at the $M_{\cal F}$ scale, nor is the subject of this work, the elevation of the light Higgs boson mass via the vector-like particle contribution is vital for all model scales.

Regarding the SUSY contribution to flavour changing neutral current processes, we evaluated the branching ratio of the rare b-quark decay, where the leading squark and gaugino contributions come in with opposite signs to that of the Standard Model and Higgs terms. A recent experimental constraint of $Br(b \to s \gamma) = (3.43 \pm 0.21^{stat}~ \pm 0.24^{th} \pm 0.07^{sys}) \times 10^{-4}$~\cite{HFAG} differs only slightly from the theoretical next-to-next-to-leading order Standard Model contribution of $Br(b \to s \gamma) = (3.15 \pm 0.23) \times 10^{-4}$~\cite{Misiak:2006zs}, and likewise, an alternative calculation of $Br(b \to s \gamma) = (2.98 \pm 0.26) \times 10^{-4}$~\cite{Becher:2006pu}. Additional uncertainties taken into account include perturbative and non-perturbative QCD corrections, which suggested a relaxation of the measured results to two standard deviations. In \fsu5, the SUSY counter terms created a sufficiently suppressing effect that kept the computed value at $Br(b \to s \gamma) = 3.5 - 3.6 \times 10^{-4}$ for M(\gl) $\ge$ 2~TeV~\cite{Li:2016bww,Ford:2019kzv}, nicely in line with the observed value. Moreover, we studied the rare process of a B-meson decay to a dimuon of $B_S^0 \to \mu^+ \mu^-$, which commences with an $(s, \bar{b})$ quark content. The Standard Model features a loop-level process leading to a $t \bar{t} \to Z^0$ event mediated by an off-shell $W$-boson. A recent Standard Model prediction is $Br(B_S^0 \to \mu^+ \mu^-) = (3.65 \pm 0.23) \times 10^{-9}$~\cite{Bobeth:2013uxa}, though measurements reveal a branching ratio of $Br(B_S^0 \to \mu^+ \mu^-) = (2.9 \pm 0.7 \pm 0.29^{th}) \times 10^{-9}$~\cite{CMS:2014xfa}, clearly placing severe constraints on the SUSY contribution. The SUSY contribution here is also adequately suppressed, such that for large gluino masses of M(\gl) $\ge$ 2~TeV, we computed branching ratios of $Br(B_S^0 \to \mu^+ \mu^-) = 3.0 - 3.2 \times 10^{-9}$~\cite{Li:2016bww,Ford:2019kzv} that conformed well with observations.

The full model space of \fsu5 shows a rather small difference $\Delta M(\widetilde{\tau}_1^{\pm}, \widetilde{\chi}_1^0)$ between the light stau \stau ~and bino LSP \lsp, generating stau-neutralino coannihilation to achieve consistency with the WMAP and Planck Collaborations~\cite{Hinshaw:2012aka,Aghanim:2018eyx} measured relic density of $\Omega_{DM} h^2 \simeq 0.12$ (for instance, see Refs.~\cite{Li:2011xua,Li:2013naa,Li:2016bww}). This is certainly true for M(\gl) $\le$ 2~TeV, however, for gluinos greater than 2~TeV, $\Delta M(\widetilde{\tau}_1^{\pm}, \widetilde{\chi}_1^0)$ steadily increases, and given the bino's conspicuously small cross-section, there is inadequate coannihilation to stabilize $\Omega_{\widetilde{\chi}_1^0} h^2$ near 0.12, and thus the dark matter density overshoots the WMAP and Planck observed value~\cite{Ford:2019kzv}. In this heavy mass region, an alternate route to $\Omega_{DM} h^2 \simeq 0.12$ became necessary and presented itself through a natural dilution factor that permits \fsu5 to overproduce thermally produced LSPs. The overproduction of LSPs in time is diluted down to the correct $\Omega_{DM} h^2$ by means of the very same master $\lambda_6$ coupling that plays a dominant role in the string derived inflationary model in Flipped $SU(5)$~\cite{Ellis:2019jha,Ellis:2019opr}\cite{Ford:2019kzv}.

The efforts to capture a scintillation induced by heavy nuclei recoiling from an elastic scattering process with a Weakly Interacting Massive Particle (WIMP) are led by the XENON100~\cite{Aprile:2018dbl}, LUX~\cite{Akerib:2016vxi}, and  PandaX-II~\cite{Tan:2016zwf} Collaborations. These direct detection experiments rely upon ultra-pure liquid xenon to establish lower bounds on the spin-independent cross-sections $\sigma_{SI}$ for scattering WIMPS from nucleons. The direct searches have been steadily probing WIMP weak-scale masses, establishing the strongest lower bound of $\sigma_{SI} \sim 5 \times 10^{-10}$~pb for around 300~GeV WIMP masses~\cite{Aprile:2018dbl}. Considering the inherently small cross-section presented by a bino LSP, which in fact dominates all \lsp ~LSPs in No-Scale \fsu5, the experiments are still more than an order of magnitude short of potentially observing evidence of scattering events of \fsu5 WIMPS with nucleons. The spin-independent cross-sections for 2.2~TeV gluinos in \fsu5 hovers near $\sigma_{SI} \sim 1 \times 10^{-11}$~pb~\cite{Li:2016bww}, which is itself only a little over an order of magnitude above the neutrino scattering floor, suggesting a fine needle to thread for the direct search teams in the coming years.

The natural two-stage unification mechanism invites a relatively $\cal{F}$-ast proton decay in \fsu5, as the GUT couplings are tightly bound at the \m32 scale. This summons a dimension six partial lifetime in the leading $(e|\mu)^+ \pi^0$ channel for a proton decay rate of $\tau_p > 10^{35}$~yrs for gluino masses M(\gl) $\gtrsim$ 2.1~TeV~\cite{Li:2016bww}. These proton lifetimes comfortably evade current detection limits formulated by Super-Kamiokande~\cite{Takhistov:2016eqm}.

\section{The Muon $g-2$ in the \bfsu5 Model}

In order to address the muon $g-2$ in No-Scale \fsu5, we must first present some details of Flipped $SU(5)$. The minimal Flipped $SU(5)$ model carries a gauge group of $SU(5)\times U(1)_{X}$, which can be embedded into the $SO(10)$ model. The only other Flipped $SU(5)$ models are two alternatives that arise from orbifold compactifications. Regarding the minimal Flipped $SU(5)$ model, the generator $U(1)_{Y'}$ in $SU(5)$ is defined as 

\bea 
T_{\rm U(1)_{Y'}}={\rm diag} \left(-\frac{1}{3}, -\frac{1}{3}, -\frac{1}{3},
 \frac{1}{2},  \frac{1}{2} \right).
\label{u1yp}
\eea
where the hypercharge is given by

\bea
Q_{Y} = \frac{1}{5} \left( Q_{X}-Q_{Y'} \right).
\label{ycharge}
\eea

The Standard Model consists of three families of fermions under $SU(5)\times U(1)_{X}$, and the quantum numbers are, respectively,

\bea
F_i={\mathbf{(10, 1)}},~ {\bar f}_i={\mathbf{(\bar 5, -3)}},~
{\bar l}_i={\mathbf{(1, 5)}},
\label{smfermions}
\eea
where $i=1, 2, 3$. We adopt the convention of $\frac{1}{2{\sqrt{10}}}$ for the normalization factor for $U(1)_X$ charges.
The Standard Model particle assignments in $F_i$, ${\bar f}_i$ and ${\bar l}_i$ are

\bea
F_i=(Q_i, D^c_i, N^c_i),~{\overline f}_i=(U^c_i, L_i),~{\overline l}_i=E^c_i~,~
\label{smparticles}
\eea
where, respectively, $Q_i$ and $L_i$ are the superfields of the left-handed quark and lepton doublets and $U^c_i$, $D^c_i$, $E^c_i$ and $N^c_i$ are the $CP$ conjugated superfields for the right-handed up-type quarks, down-type quarks, leptons and neutrinos. In order to generate the heavy right-handed neutrino masses, three Standard Model singlets $\phi_i$ are included.

The GUT and electroweak gauge symmetries can be broken by two pairs of Higgs representations:

\begin{eqnarray}
H&=&{\mathbf{(10, 1)}},~{\overline{H}}={\mathbf{({\overline{10}}, -1)}}, \nonumber \\
h&=&{\mathbf{(5, -2)}},~{\overline h}={\mathbf{({\bar {5}}, 2)}}.
\label{Higgse1}
\end{eqnarray}
The $H$ multiplet states are annotated in the same manner as the $F$ multiplet in addition to a ``bar'' added above the fields for ${\overline H}$. Specifically, the Higgs particles are

\bea
H=(Q_H, D_H^c, N_H^c)~,~
{\overline{H}}= ({\overline{Q}}_{\overline{H}}, {\overline{D}}^c_{\overline{H}}, 
{\overline {N}}^c_{\overline H})~,~\,
\label{Higgse2}
\eea
\bea
h=(D_h, D_h, D_h, H_d)~,~
{\overline h}=({\overline {D}}_{\overline h}, {\overline {D}}_{\overline h},
{\overline {D}}_{\overline h}, H_u)~,~\,
\label{Higgse3}
\eea
and note that $H_d$ and $H_u$ are one pair of Higgs doublets in the MSSM. Furthermore, one Standard Model singlet $\Phi$ is admitted.

To break the $SU(5)\times U(1)_{X}$ gauge symmetry down to the Standard Model gauge symmetry, we insert the following Higgs superpotential at the GUT scale:

\bea
{\it W}_{\rm GUT}=\lambda_1 H H h + \lambda_2 {\overline H} {\overline H} {\overline
h} + \Phi ({\overline H} H-M_{\rm H}^2)~.~ 
\label{spgut} 
\eea
The superpotential consists of only one F-flat and D-flat direction, though this can always be rotated along the $N^c_H$ and ${\overline {N}}^c_{\overline H}$ directions. In that event, we obtain $<N^c_H>=<{\overline {N}}^c_{\overline H}>=M_{\rm H}$. Through the supersymmetric Higgs mechanism, the superfields $H$ and ${\overline H}$ can be ``eaten'' and gain large masses, except for $D_H^c$ and ${\overline {D}}^c_{\overline H}$. Additionally, the superpotential terms $ \lambda_1 H H h$ and $ \lambda_2 {\overline H} {\overline H} {\overline h}$ couple the $D_H^c$ and ${\overline {D}}^c_{\overline H}$ with the $D_h$ and ${\overline {D}}_{\overline h}$, respectively, the result of which is the formation of massive eigenstates with masses $2 \lambda_1 <N_H^c>$ and $2 \lambda_2 <{\overline {N}}^c_{\overline H}>$. A significant consequence of this is that the doublet-triplet splitting due to the missing partner mechanism proceeds naturally. The triplets in $h$ and ${\overline h}$ maintain only small mixing through the $\mu$ term, therefore, the Higgsino-exchange mediated proton decay is negligible, and as a result, for example, the dimension-5 proton decay problem is absent. 

As we discussed, string-scale gauge coupling unification can be realized by introducing these two sets of TeV-scale vector-like multiplets~\cite{Jiang:2006hf}:
\begin{eqnarray}
&& XF ={\mathbf{(10, 1)}}~,~{\overline{XF}}={\mathbf{({\overline{10}}, -1)}}~,~\\
&& Xl={\mathbf{(1, -5)}}~,~{\overline{Xl}}={\mathbf{(1, 5)}}~.~\,
\end{eqnarray}
The particle content of the multiplets under Standard Model gauge symmetry can be identified through decompositions of $XF$, ${\overline{XF}}$, $Xl$, and ${\overline{Xl}}$: 
\begin{eqnarray}
&& XF = (XQ, XD^c, XN^c)~,~ {\overline{XF}}=(XQ^c, XD, XN)~,~~~~~\\
&& Xl= XE~,~ {\overline{Xl}}= XE^c~.~
\end{eqnarray}
Finally, the quantum numbers under the $SU(3)_C \times SU(2)_L \times U(1)_Y$ gauge symmetry for the vector-like multiplets are 
\begin{eqnarray}
&& XQ={\mathbf{(3, 2, \frac{1}{6})}}~,~
XQ^c={\mathbf{({\bar 3}, 2,-\frac{1}{6})}} ~,~\\
&& XD={\mathbf{({3},1, -\frac{1}{3})}}~,~
XD^c={\mathbf{({\bar 3},  1, \frac{1}{3})}}~,~\\
&& XN={\mathbf{({1},  1, {0})}}~,~
XN^c={\mathbf{({1},  1, {0})}} ~,~\\
&& XE={\mathbf{({1},  1, {-1})}}~,~
XE^c={\mathbf{({1},  1, {1})}}~.~\,
\label{qnum1}
\end{eqnarray}

\begin{figure}[t]
   \centering
        \includegraphics[width=0.4\textwidth]{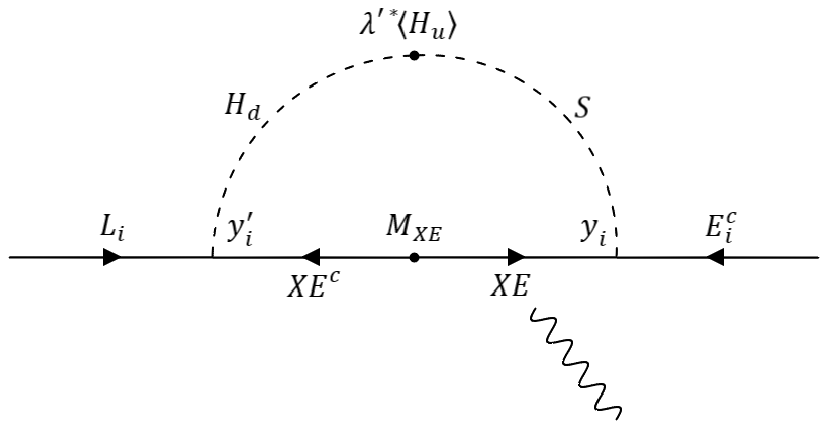}
        \caption{Chirality flip Feynman diagram depicting the inclusion of the singlet $S$ that couples with the $(XE,XE^c)$ of our $Xl$ and $\overline{Xl}$ vector-like flippon multiplet sets.}
        \label{fig:fdiagram}
\end{figure}

We now introduce a Standard Model singlet $S$~\cite{Lindner:2016bgg, Calibbi:2018rzv} that can explain within No-Scale \fsu5 the BNL+FNAL muon anomalous magnetic moment results. This presents us with the following superpotential:
\begin{eqnarray}
W &=& \lambda' S H_d H_u + y_i S XE E_i^c + y'_i L_i XE^c H_d \nonumber \\
&& + M_{XE} XE^c XE~.~\,
\label{qnum2}
\end{eqnarray}
The inclusion of the singlet $S$ allows the muon anomalous magnetic moment to be easily explained, though in No-Scale \fsu5 we can directly exploit the already present $(XE,XE^c)$ of our $Xl$ and $\overline{Xl}$ vector-like multiplet sets. The Feynman diagram is presented in FIG.~\ref{fig:fdiagram}, which provides a positive contribution to $\Delta a_{\mu}$. In general, $S$ will acquire a VEV, and in this regard, No-Scale \fsu5 is then similar to the Next to the Minimal Supersymmetric Standard Model (NMSSM). In addition, we can introduce a tadpole term $\Lambda^2 S$ into the above superpotential. Given some fine-tuning, we can obtain $\langle S \rangle =0 $ by choosing $\Lambda^2 + \lambda' \langle H_d \rangle \langle H_u \rangle =0$. Consequently, the discussions for the Higgs potential and Higgs mass are the same as our previous studies.

Due to the chirality flip in the Feynman digram, it is well known that the vector-like particle contribution will be large enough to handily accommodate the observed BNL+FNAL measurements. Nonetheless, we can provide here a rough estimate of the contributions as a means of illustrating that it is indeed sufficient. The contribution from the vector-like particles can be approximated by $\Delta a_{\mu} \sim m_{\mu} v / M^2$~\cite{Calibbi:2018rzv}, where $m_{\mu}$ is the mass of the muon, $v$ is the VEV that breaks $SU(2)_L$, and $M^2$ is the scale of the fields, in our case the charged vector-like ``lepton''/singlet $(XE, XE^c)$. We shall use the up VEV in No-Scale \fsu5, which we know from precise spectra calculations is $v_u \approx 177$~GeV. For $M^2$ we can apply the vector-like mass scale $M_V$, which for $M(\widetilde{g}) \approx 2.2$~TeV and $m_t = 173.3$~GeV is $M_V \approx 90$~TeV~\cite{Li:2016bww}. These give $\Delta a_{\mu}(XE) \sim 23.1 \times 10^{-10}$, which when summed with the SUSY contribution for these same gluino and top masses of $\Delta a_{\mu}(\rm SUSY) \approx 2.1 \times 10^{-10}$~\cite{Li:2016bww}, generates a total of $\Delta a_{\mu}({\rm SUSY}+XE) \sim 25.2 \times 10^{-10}$, compared to the observed value of $25.1 \times 10^{-10}$. Realize that this is a streamlined evaluation and we did not intend for this brief computation to strike the bullseye, though it does effectively demonstrate that our methodology is sound and can resolve the anomalous BNL+FNAL findings.

\section{LHC Constraints}

The LHC Run 2 has accumulated up to 139 \fb of collision data through 2020, yet, no indications of new physics have been observed. We now update the No-Scale \fsu5 model with respect to the latest LHC Run 2 data set. Our primary SUSY interest here lies with the gluino \gl, light stop \ls, light chargino \char (= \neu2), and light sleptons \stau and \sel. The ATLAS and CMS Collaborations study a myriad of simplified model scenarios, but No-Scale \fsu5 is a realistic physical model, and as such possesses very defined cascade decay chains to final states. We thus only consider those few simplified models applicable to No-Scale \fsu5. 

\begin{figure*}[t]
   \centering
     \begin{tabular}{cc}
        \includegraphics[width=0.40\textwidth]{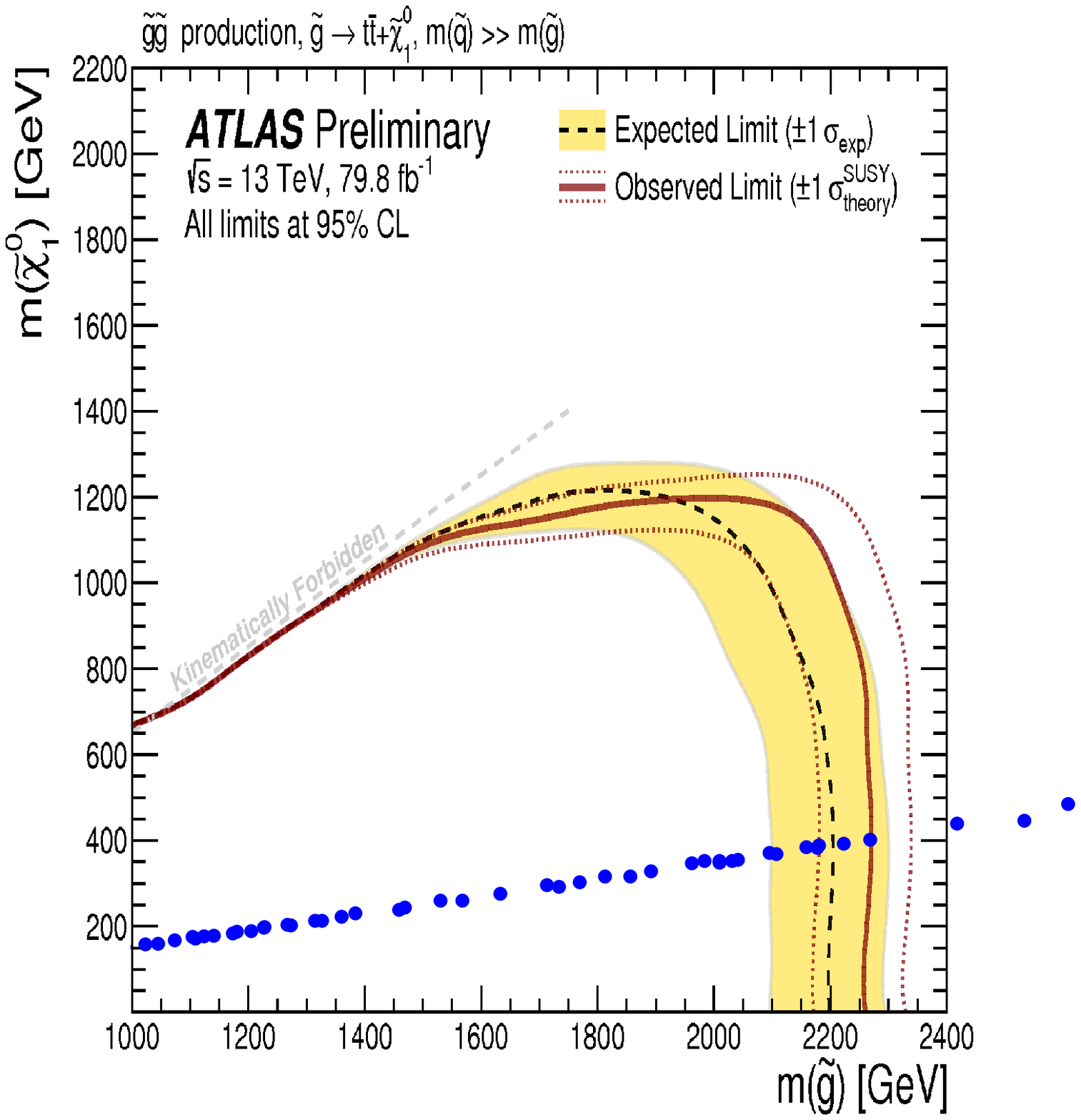}
        \includegraphics[width=0.40\textwidth]{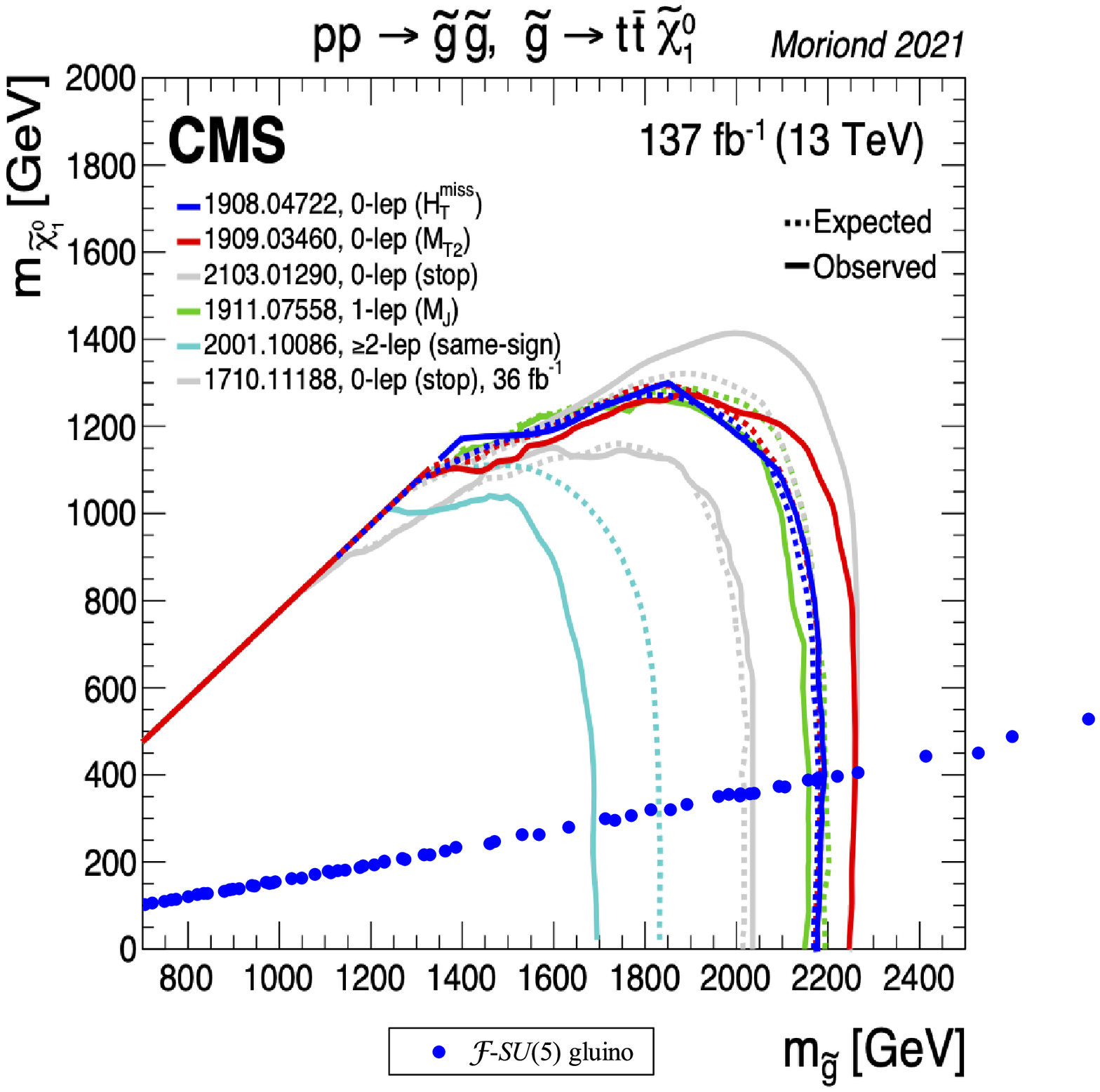} \\
        \includegraphics[width=0.40\textwidth]{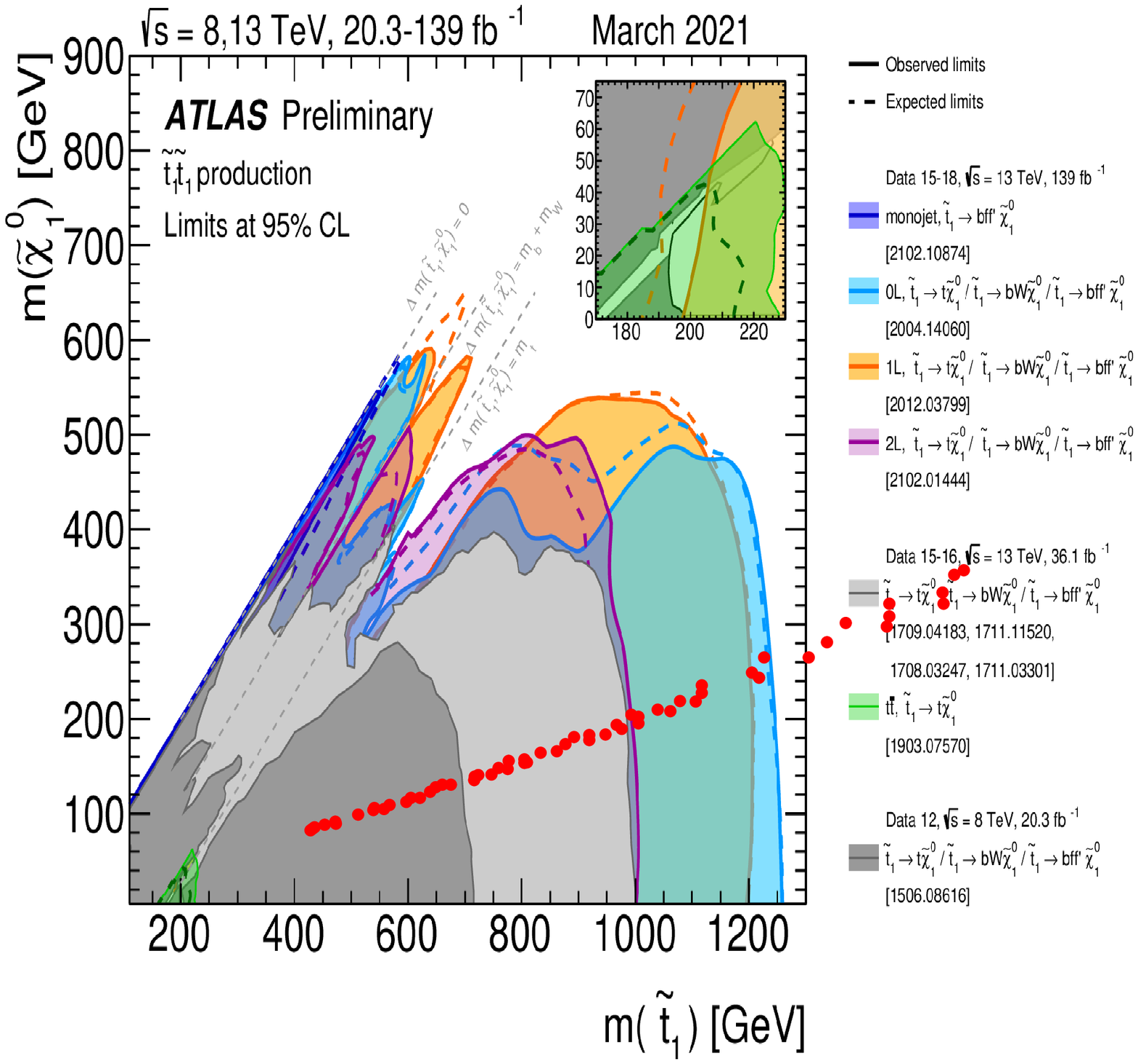}
        \includegraphics[width=0.40\textwidth]{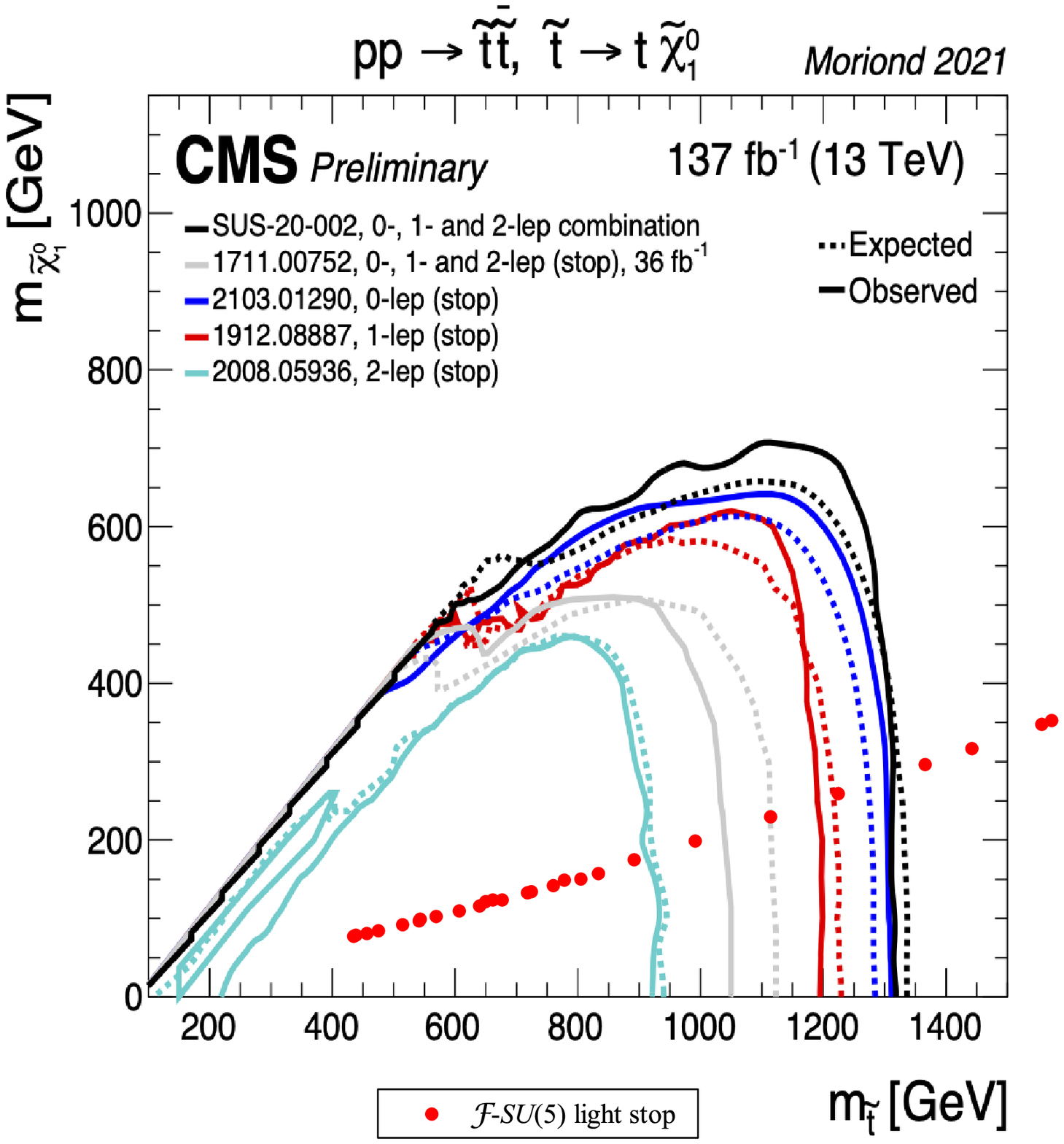} \\
        \includegraphics[width=0.40\textwidth]{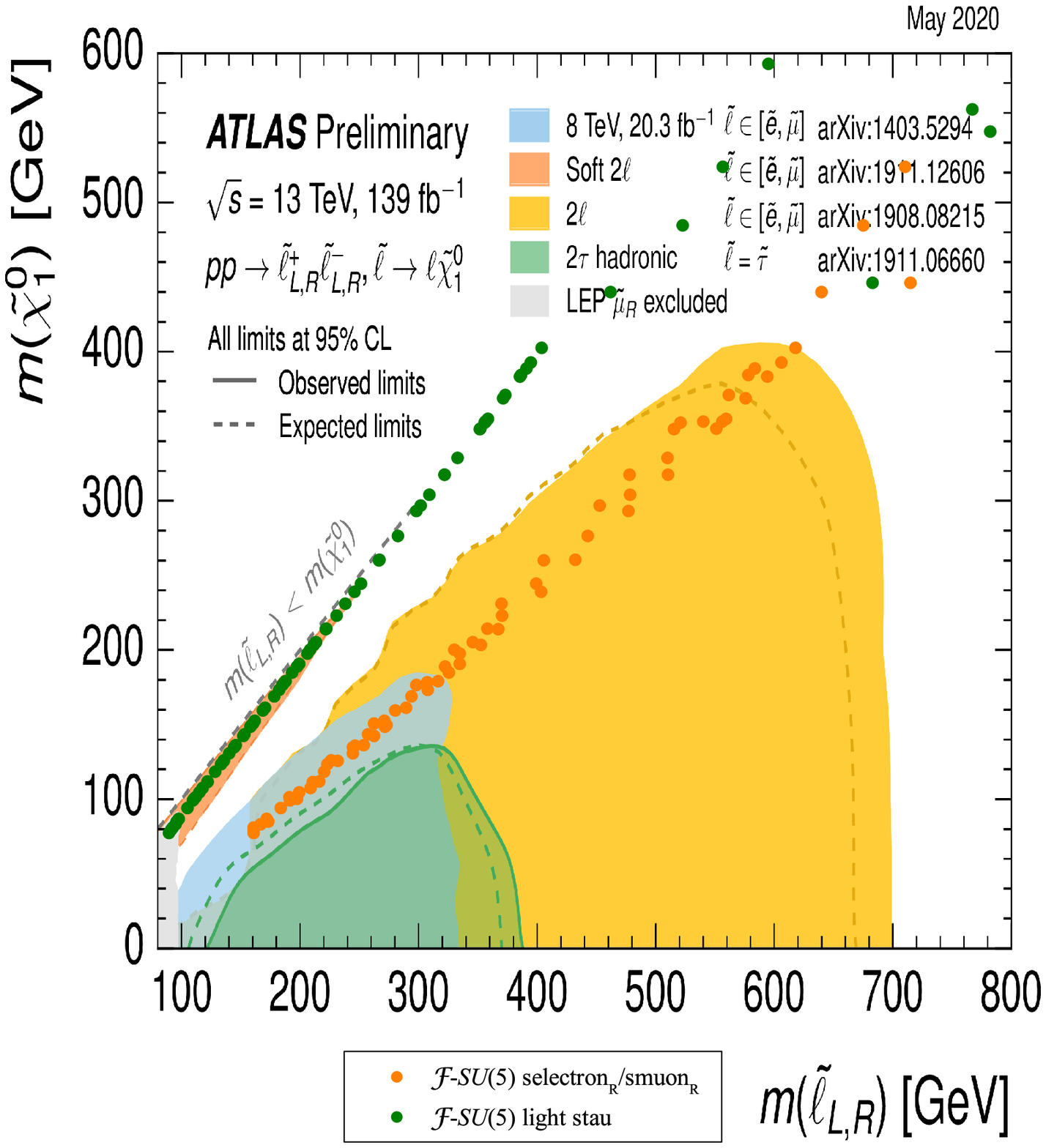}
        \includegraphics[width=0.40\textwidth]{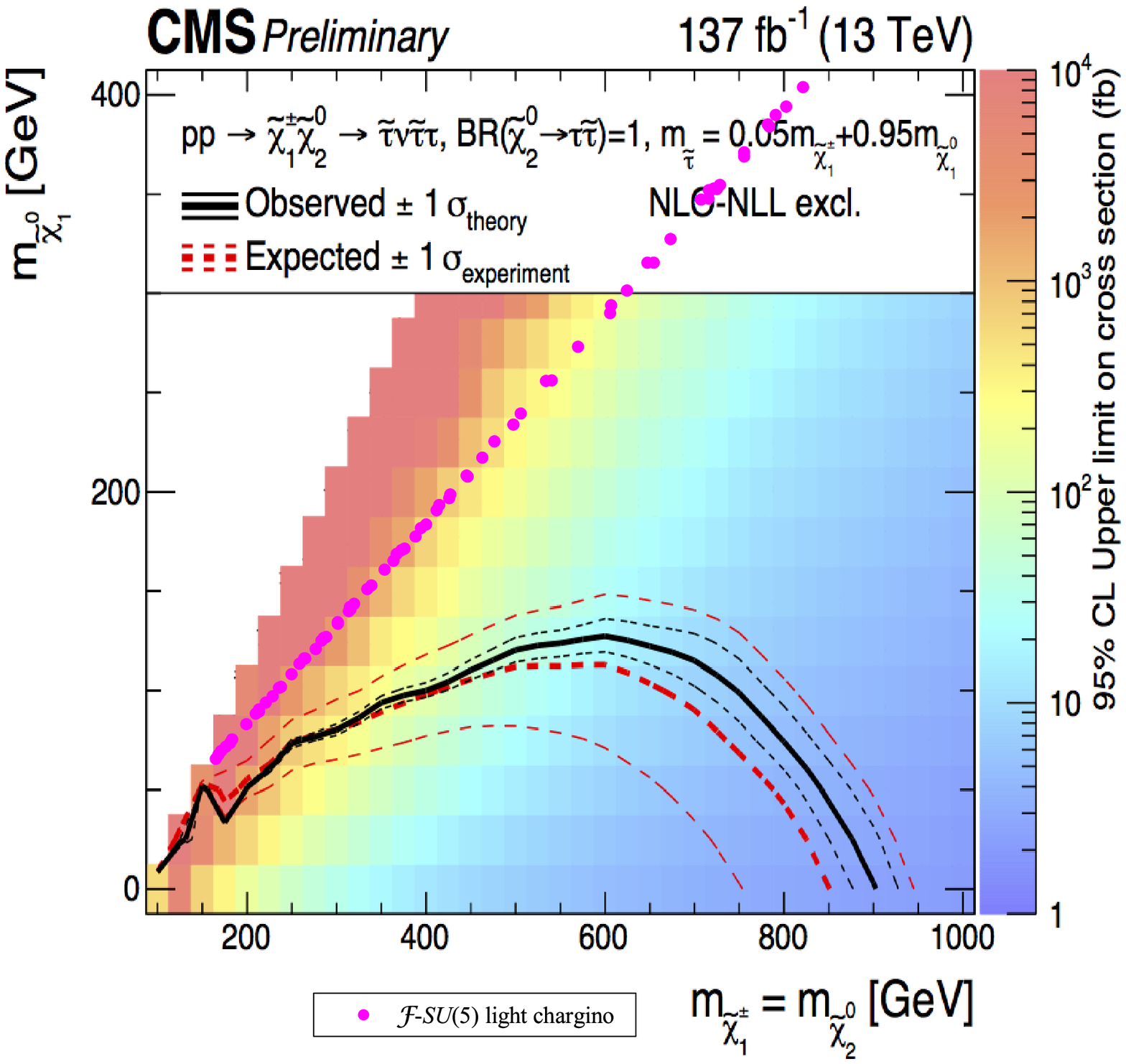} \\
    \end{tabular}
        \caption{Illustration of No-Scale \fsu5 consistency with the most recent LHC constraints on gluino, light stop, slepton, and chargino masses. In this Figure the model space is superimposed on top of the ATLAS~\cite{ATLAS:2018yhd,Aad:2015pfx,Aaboud:2017nfd,Aaboud:2017ayj,Aaboud:2017phn,Aaboud:2017aeu,Aaboud:2019hwz,Aad:2020sgw,Aad:2020aob,Aad:2021hjy,Aad:2021egl,Aad:2014vma,Aad:2019vnb,Aad:2019byo,Aad:2019qnd} and CMS~\cite{Sirunyan:2017pjw,Sirunyan:2019ctn,Sirunyan:2019xwh,CMS:2019tlp,Sirunyan:2020ztc,Sirunyan:2017leh,Sirunyan:2019glc,CMS:2021qkg,Sirunyan:2020tyy,Sirunyan:2021mrs,CMS:2021bra} exclusion plots. The model is represented by the small dots which are comprised of the prior benchmark sets of Refs.~\cite{Li:2011xua,Li:2011in,Li:2011gh,Li:2013naa,Li:2013mwa,Li:2016bww,Ford:2019kzv}. The No-Scale \fsu5 model space runs well beyond the right side margin of the LHC plots, as large as M(\gl) $\sim$ 7.5~TeV, an energy well beyond the reach of the LHC Run2. These figures are evidence that No-Scale \fsu5 remains viable at the LHC, and absent any statistically significant discovery in the near future, will be probed at the LHC and/or future higher energy colliders for many years to come.}
        \label{fig:lhc}
\end{figure*}

The model is evaluated against the Run 2 data in FIG.~\ref{fig:lhc}. We superimpose the model points onto the ATLAS~\cite{ATLAS:2018yhd,Aad:2015pfx,Aaboud:2017nfd,Aaboud:2017ayj,Aaboud:2017phn,Aaboud:2017aeu,Aaboud:2019hwz,Aad:2020sgw,Aad:2020aob,Aad:2021hjy,Aad:2021egl,Aad:2014vma,Aad:2019vnb,Aad:2019byo,Aad:2019qnd} and CMS~\cite{Sirunyan:2017pjw,Sirunyan:2019ctn,Sirunyan:2019xwh,CMS:2019tlp,Sirunyan:2020ztc,Sirunyan:2017leh,Sirunyan:2019glc,CMS:2021qkg,Sirunyan:2020tyy,Sirunyan:2021mrs,CMS:2021bra} exclusion plots for easy visual recognition of the model's status. For points we use all the prior benchmark sets published in Refs.~\cite{Li:2011xua,Li:2011in,Li:2011gh,Li:2013naa,Li:2013mwa,Li:2016bww,Ford:2019kzv}, which are represented by the small dots in FIG.~\ref{fig:lhc}. Note that No-Scale \fsu5 model points effectively run up to $M$(\gl) $\sim$ 7.5~TeV~\cite{Ford:2019kzv}, well beyond the display region of the FIG.~\ref{fig:lhc} plot spaces and the reach of the LHC Run 2. {\it It is significant that only the low mass region of the model has been excluded to date, with the majority of the model yet to be probed at the LHC and/or future higher energy colliders in the years to come}.

Given the fact the $M_3$ gluino runs flat from the lower stage unification scale \m32 at about $10^{16}$~GeV down to the vector-like multiplet decoupling scale \mv, this prevents the gluino mass $M$(\gl) from running at 2-loops beyond all the squark masses. Consequently, $M$(\gl) is lighter than all squarks except the light stop, generating the distinctive SUSY mass signature of $M(\widetilde{t}_1) < M(\widetilde{g}) < M(\widetilde{q})$, where here we identify $\widetilde{q} = \{ \widetilde{u}, \widetilde{d}, \widetilde{c}, \widetilde{s}, \widetilde{b}, \widetilde{t}_2 \}$. This characteristic descending of the gluino due the vector-like particles leaves only one available channel, \gl $\to$ \ls t, at 100\% for spectra with $M$(\gl) $\gtrsim$ 1~TeV, with \ls $\to t \bar{t}$+\lsp ~also at 100\% in the same region. Therefore, the ATLAS and CMS simplified models for the gluino and light stop shown in FIG.~\ref{fig:lhc} are descriptive of No-Scale \fsu5 decay channels. Below $M$(\gl) $\sim$ 1~TeV, the gluino and light stop become compressed, resulting in an either an off-shell top quark or light stop, though we shall not discuss that more complicated option since that region of gluino mass has already been excluded by both LHC Run1 and Run 2 results.

Furthermore, in the region of large gluino in No-Scale \fsu5, electroweak production of charginos, neutralinos, and sleptons leads to the subsequent decays of \char $\to$ \stau $\nu_{\tau}$, \neu2 $\to$ \stau $\tau^{\mp}$, \stau $\to$ $\tau^{\pm}$ + \lsp, and \sel $\to$ $e^-$ + \lsp, with branching ratios at either 100\% or near 100\% for all four processes. The closest fit to \fsu5 for \char \neu2 pair production is the CMS simplified model \char \neu2 $\to \widetilde{\tau} \nu \widetilde{\tau} \tau$, BR(\neu2 $\to \tau \widetilde{\tau}$) = 1, $m_{\widetilde{\tau}} = 0.05m_{\widetilde{\chi}^{\pm}_1} + 0.95m_{\widetilde{\chi}^0_1}$, and therefore, the sole exclusion curves applicable to \fsu5 for \char \neu2 production, as depicted in FIG.~\ref{fig:lhc}.

\section{Conclusions}

A common utterance throughout history has always been ``necessity is the mother of invention''. Stated analogously, new discoveries are pursued by new ideas. The FNAL measured confirmation of the BNL $g-2$ of the muon discrepancy with the theoretical Standard Model value compels an argument for the origins of the 4.2$\sigma$ deviation, serving as a call for any and all rationale regarding this observed discordance amongst theory and experiment. Given the eventual likelihood of a 5$\sigma$ discovery, any serious GUT candidate must be naturally equipped to elegantly assimilate new physics into the muon's magnetic moment.

The legitimate natural GUT candidate explored in this paper is No-Scale \fsu5, which merges the top-down theoretical dynamics of No-Scale Supergravity with the Flipped $SU(5)$ GUT and additional string derived vector-like matter. The extra TeV-scale $10 + \overline{10}$ vector-like multiplet and charged vector-like singlet $(XE,XE^c)$, referred to as flippons, are pivotal here in that the charged vector-like ``lepton''/singlet couples with the muon, an added singlet $S$, and the supersymmetric down-type Higgs $H_d$, to handily account for the positive contribution to the $g-2$ of the muon via a chirality flip.

Finally, it should be noted how fascinating it is that the same vector-like multiplets that were introduced to elevate the light Higgs boson mass to its observed 125 GeV value now stage Act Two, by likewise raising the muonic $g-2$ to its measured range. While we cannot conclusively purport that this is not a mere coincidence, it is surely nonetheless intriguing given the deep fundamental alliance this GUT model holds with superstring theory and cosmology.


\section{Acknowledgments}

Portions of this research were conducted with high performance computational resources provided 
by the Louisiana Optical Network Infrastructure (http://www.loni.org). This research was supported 
in part by the Projects 11475238, 11647601, and 11875062 supported 
by the National Natural Science Foundation of China (TL), 
by the Key Research Program of Frontier Science, Chinese Academy of Sciences (TL),
and by the DOE grant DE-FG02-13ER42020 (DVN). 


\bibliography{bibliography}

\end{document}